\pdfoutput=1
\documentclass[10pt]{article}
\usepackage{graphicx}

\def\Title#1{\begin{center} {\Large #1 } \end{center}}
\def\Author#1{\begin{center}{ \sc #1} \end{center}}
\def\Address#1{\begin{center}{ \it #1} \end{center}}

\newcommand\pubblock{\rightline{\begin{tabular}{l} Proceedings of the Fifth Annual LHCP\\ \pubnumber\\
         \pubdate  \end{tabular}}}

\newenvironment{Abstract}{\begin{quotation} \begin{center} 
             \large ABSTRACT \end{center}\bigskip 
      \begin{center}\begin{large}}{\end{large}\end{center} \end{quotation}}

\newenvironment{Presented}{\begin{quotation} \begin{center} 
             PRESENTED AT\end{center}\bigskip 
      \begin{center}\begin{large}}{\end{large}\end{center} \end{quotation}}

\def\beq{\begin{equation}}
\def\eeq#1{\label{#1}\end{equation}}
\def\eeqn{\end{equation}}

\def\beqa{\begin{eqnarray}}
\def\eeqa#1{\label{#1}\end{eqnarray}}
\def\eeqan{\end{eqnarray}}

\let\bar=\overbar

\def\Dslash{\not{\hbox{\kern-4pt $D$}}}
\def\dslash{\not{\hbox{\kern-2pt $\del$}}}

\def\msb{{\bar{\ssstyle M \kern -1pt S}}}

\usepackage{color}
\usepackage{amsmath}
\usepackage{bm}
\usepackage{textcomp}
\newcommand\pubnumber{ }

\newcommand\pubdate{\today}

\def\affiliation{
On behalf of the ATLAS, CMS and LHCb Experiments, \\
European Organization for Nuclear Research (CERN), \\
Route de Meyrin 385, 1217 Meyrin, Switzerland}

\begin{document}

% large size for the first page
\large
\begin{titlepage}
\pubblock

%% Change the title, name, abstract
%% Title 
\vfill
\Title{$C\!P$ VIOLATION IN BEAUTY AND CHARM }
\vfill

%  if you need to add the support use this, fill the \support definition above. 
%   \Author{ FIRSTNAME LASTNAME \support }
\Author{ DORDEI FRANCESCA  }
\Address{\affiliation}
\vfill
\begin{Abstract}
The study of $C\!P$ violation in decays of beauty- and charm-hadrons provides a fundamental test of the predictions of the Standard Model (SM) and represents a sensitive probe to search for physics effects that cannot be described within the SM. In these proceedings, precision measurements are presented of several observables related to the so-called unitary triangles, that arise from the unitarity requirements on the 3x3 Cabibbo-Kobayashi-Maskawa matrix describing the quark mixing. Moreover, recent measurements of $C\!P$ violation in beauty- and charm-hadrons are illustrated. In particular, a recent analysis providing the first evidence for $C\!P$ violation in the beauty baryon sector and the world best precise measurement of $C\!P$ violation in the charm system will be described in detail. The discussed analyses are based on proton-proton collision data collected during 2011 and 2012 (LHC Run 1) by the ATLAS, CMS and LHCb experiments. 
\end{Abstract}
\vfill

% DO NOT CHANGE 
\begin{Presented}
The Fifth Annual Conference\\
 on Large Hadron Collider Physics \\
Shanghai Jiao Tong University, Shanghai, China\\ 
May 15-20, 2017
\end{Presented}
\vfill
\end{titlepage}
\def\thefootnote{\fnsymbol{footnote}}
\setcounter{footnote}{0}
%

% normal size for the rest
\normalsize 

%% Your paper should be entered below. 

\section{Introduction}

Although the Standard Model (SM) has been very successful in describing existing data~\cite{PDG}, it is well known that it must be an effective theory whose validity breaks down at some higher energy scale (larger than few hundred GeV)~\cite{Ellis}. For example, contributions from physics Beyond the Standard Model (BSM) are needed~\cite{baryon2} in order to explain the large matter-antimatter asymmetry that resulted in the matter dominated universe we observe today~\cite{baryon}. In the SM, the only source of $C\!P$ violation is a complex phase in the Cabbibo-Kobayashi-Maskawa (CKM) quark mixing matrix~\cite{CKM} which is many orders of magnitude too small to match the observed matter-antimatter asymmetry~\cite{gavela}. The CKM matrix is a 3x3 matrix that parametrises transitions between the three quark families due to charged current interactions. It is fully described by four independent real parameters,  namely  three  rotation  angles  and  one  complex  phase, that are not predicted by the SM but need to be measured. A  non-zero  value  of the  complex  phase  can  induce $C\!P$ violating  asymmetries  in  processes  in  which  two  or  more amplitudes  with  different  weak  phases  interfere.   

Most of the BSM physics models available today predict the existence of additional unobserved heavy particles that can appear in internal lines of higher order diagrams contributing to a process (so called box and penguin diagrams). Those particles should produce sizeable modifications of observables related to these processes, such as $C\!P$ violating parameters. Therefore, precision measurements of $C\!P$ violation in the beauty and charm systems are mandatory in order to confirm SM predictions and to check whether non-SM contributions are also present. Since these new unobserved particles appear only as virtual particles in internal lines of loop diagrams, indirect searches for BSM physics are sensitive to much higher mass scales  than direct searches. The reachable mass scale can indeed exceed the centre of mass energy of the accelerator at which the measurement
is performed, reaching up to $\mathcal{O}$(10-100 TeV).\\

There are three categories of $C\!P$ violation which are introduced in the following in order to better understand the results of the measurements presented. 

The first type of $C\!P$ violation is possible for both charged and neutral meson 
decays
\begin{enumerate}
 \item \textbf{$C\!P$ violation in decay}. It occurs when the amplitude for a decay and its $C\!P$ conjugate decay have different magnitudes. If $A_f$ and $\overline{A}_{\overline{f}}$ are 
the amplitudes for the process $B \to f$ and $\overline{B} \to \overline{f}$, respectively, where $f$ and $\overline{f}$ are $C\!P$-conjugate multi-particle final 
states, $C\!P$ violation in decay occurs if 
\begin{equation}\nonumber
  \left| A_f \right| \neq \left| \overline{A}_{\overline{f}} \right| \,.
  \end{equation}
\end{enumerate}

The other two categories are only possible for neutral mesons, hereafter referred to as $P^0$, which exhibit the phenomenon of mixing. Since the weak interaction eigenstates are not the same as the mass eigenstates, the flavour content of neutral mesons changes as a function of time, giving rise to the phenomenon known as neutral meson mixing. The phenomenological aspects linked to neutral meson mixing are described in many articles, e.g.~\cite{Nierste}. Only the main parameters involved are briefly introduced here. Any arbitrary combination of flavour eigenstates, $P^0$ and $\overline{P}^0$, has a time evolution described by an effective Schr{\"o}dinger equation (units are such that $\hbar=\textit{c}=1$)
\begin{equation}\label{eq:Scheq}
\textit{i}\,\,\frac{d}{dt}\begin{pmatrix} P^0 \\ \overline{P}^0 \\ \end{pmatrix} = 
 \begin{pmatrix}
  M_{11}-i\,\Gamma_{11}/2 & M_{12}-i\,\Gamma_{12}/2  \\
  M_{21}-i\,\Gamma_{21}/2 & M_{22}-i\,\Gamma_{22}/2  \\
 \end{pmatrix} \begin{pmatrix} P^0 \\ \overline{P}^0 \\ \end{pmatrix}\,,
\end{equation}
where \textbf{M} and $\bm{\Gamma}$ are $2\times2$ hermitian matrices. The heavy (H) and light (L) mass eigenstates of the Schr{\"o}dinger equation are obtained diagonalizing the matrix in Eq.\ref{eq:Scheq}
\begin{equation}
 |P_{H/L}\rangle = p |{P}^0\rangle \pm q |\overline{P}^0\rangle \,,
\end{equation}
where the complex coefficients p and q obey the normalization condition $|p|^2+|q|^2=1$. The system can be defined in terms of four different experimental observables. They are the mass difference $\Delta m$ and the decay width difference $\Delta \Gamma$ between the mass eigenstates, defined by
\begin{equation}
\Delta m \equiv m_H - m_L \simeq 2|M_{12}|, \,\,\,\,\,\,\,\,\,\,\,\,\,\,\,\,\,\,\,\,\, \Delta \Gamma \equiv \Gamma_L - \Gamma_H \simeq 2 |\Gamma_{12}| \cos(\phi_{12}),
\end{equation}
and the average mass and width, defined by
\begin{equation}
m = \frac{m_H+m_L}{2}, \,\,\,\,\,\,\,\,\,\,\,\,\,\,\,\,\,\,\,\,\, \Gamma = \frac{\Gamma_L+\Gamma_H}{2}.
\end{equation}

Thanks to the phenomenon of mixing, neutral mesons are an ideal laboratory to test the other two categories of $C\!P$ violation

\begin{enumerate}
\setcounter{enumi}{1}
\item \textbf{$C\!P$ violation in mixing}. It occurs if the 
amplitude of 
the process $P^0 \to \overline{P}^0$ is different with respect to the opposite process, $\overline{P}^0 \to P^0$, i.e. $|q/p|$ not equal to 1. It does not depend on the final state. 

\item \textbf{$C\!P$ violation in the interference between mixing and decay}. If there is a final state, $f$, 
common to $P^0$ and $\overline{P}^0$ (e.g. the final state is a $C\!P$ eigenstate) the neutral meson can either decay to the final state directly or first oscillate and then decay. This gives rise to $C\!P$ violation in the interference between
mixing and decay. It occurs if
\begin{eqnarray}
\mathcal{I}m(\lambda_f) & \equiv & \mathcal{I}m \left( \frac{q}{p}\, \frac{\overline{A}_f}{A_f} \right) \neq 0
\,    , \label{deflaf}
\end{eqnarray}
where $\lambda_f$ is a parameter that quantifies this kind of $C\!P$ violation.
Assuming $C\!P\!T$ invariance, the $C\!P$ asymmetry as a function of time for neutral \textit{P} mesons decaying to a $C\!P$ eigenstate $f$ is given by
\begin{equation}
A^{\mathrm{C\!P}}(t) =\frac{\Gamma_{\overline{P}^0\to f}(t) - \Gamma_{P^0 \to f}(t)}{\Gamma_{\overline{P}^0\to f}(t) +  \Gamma_{P^0 \to f}(t)} = \frac{-C_{f} \cos(\Delta mt) + S_{f} \sin(\Delta m t)}{\cosh \left( \frac{\Delta \Gamma }{2}t \right) - \mathcal{A}^{\Delta \Gamma }_{f} \sinh \left( \frac{\Delta \Gamma }{2}t \right)}\,,
\end{equation}
The quantities $C_f$, $S_f$, and $\mathcal{A}^{\Delta \Gamma_{(s)}}_f$ are
\begin{equation}\label{adeltagamma}
  C_f = \frac{1-|\lambda_f|^2}{1+|\lambda_f|^2}, \,\,\,\,\,\,\,\,\,\,\,\, S_f = \frac{2 \mathrm{Im} \lambda_f}{1+|\lambda_f|^2}\,,  \,\,\,\,\,\,\,\,\,\,\,\, \mathrm{and}\,\,\,\,\,\,\,\,\,\,\,\, \mathcal{A}^{\Delta \Gamma_{(s)}}_f=-\frac{2\mathrm{Re}\lambda_f}{1+|\lambda_f|^2}\,,
\end{equation}
where the parameter $C_f$ and $S_f$ are related to $C\!P$ violation in decay and $C\!P$ violation in the interference between mixing and decay, respectively.
The parameter $\mathcal{A}^{\Delta \Gamma_{(s)}}_f$ can be expressed as \mbox{$\mathcal{A}^{\Delta \Gamma}_f = \pm \sqrt{1-C_f^2-S_f^2}$}.
\end{enumerate}	

\indent In the following a selection of recent ATLAS, CMS and LHCb measurements of $C\!P$ violation are presented. They are based on 2011+2012 data sets. ATLAS~\cite{ATLAS}, CMS~\cite{CMS} and LHCb~\cite{LHCb} are three of the four large particle experiments located at the Large Hadron Collider (LHC). ATLAS and CMS are general purpose detectors which cover the full spectrum of high-energy physics, while LHCb is specialised for precision measurements of $b$- and $c$- hadron decays.
 
\section{CKM metrology}
The  unitarity  of  the  CKM  matrix  gives  rise  to  six  orthogonality  relations,  which  can  be
represented as triangles in the complex plane. The triangle derived from multiplying the first and
third columns of the CKM matrix, $V_{ud}V_{ub}^*+V_{cd}V_{cb}^*+V_{td}V_{tb}^*=0$, commonly referred to as the $B^0$ triangle, is of particular interest since its three sides are of comparable
length. Its three angles are closely related to $C\!P$ violating observables and the lengths of its sides can be measured from
$C\!P$ conserving observables. A similar triangle can be also drawn for the $B^0_s$ system. The overconstrained determination of the sides and angles of these Unitarity Triangles provides a powerful consistency test of the SM picture of $C\!P$ violation. 

\subsection{Measurement of the CKM phase $\gamma$}
\label{subsec:gamma}
The  least  well  determined  parameter  of  the  Unitarity  Triangle  to  date  is  the  angle $\gamma \equiv - \mathrm{arg}(V_{ud}V_{ub}^*/V_{cd}V_{cb}^*)$. This angle provide a standard candle for the SM since it is the only one that can be determined from tree-level only processes, thus it is expected to be unaffected by possible contributions from BSM. Comparisons between direct measurements of $\gamma$ and constraints coming from global fits, which include higher order diagrams as well, represent a very stringent test of the SM. In order to reach this goal it is mandatory to improve the precision on tree-level measurements of $\gamma$ since the indirect constraints are quite precise. Indeed, from a global fit of different CKM parameters the CKMfitter group derives $\gamma=\left( 65.3^{+1.0}_{-2.5} \right)^{\circ}$~\cite{Charles:2015gya}. Experimentally, this angle is measured exploiting the interference between $b \to c$ and $b \to u$ transitions. 
By building the $C\!P$ asymmetry between $B^\mp \to D^0 K^\mp \to f_D K^\mp$ and $B^\mp \to \overline{D}^0 K^\mp \to f_D K^\mp$ decays, where $f_D$ is a final state common to both $D^0$ and $\overline{D}^0$, it is possible to determine the value of $\gamma$. However, the measurements are experimentally challenging since they involve purely hadronic
final  states,  which  are  difficult  to  trigger,  and  require  excellent $K/\pi$ separation.   Moreover,
branching fractions for the interesting decays are small. Thus, the combination of a plethora of independent modes is the only way to achieve the ultimate precision. Depending on the $D^0$ final state $f_D$, three main methods can be distinguished: in the ADS method $C\!P$ eigenstates like $D^0 \to K^+ \pi^-$ are used, in the GLW method $C\!P$-specific $D^0$ decays like $D^0 \to K^+ K^-$ or $D^0 \to \pi^+ \pi^-$ are used, and in the GGSZ method multi-body $D^0$ decays are exploited, which require the analysis of the corresponding Dalitz plot. Another independent approach is to measure time-dependent $C\!P$ asymmetries in the decay $B^0_s \to D_s^+ K^-$.\\
\begin{figure}[h]
\centering
\includegraphics[height=2in]{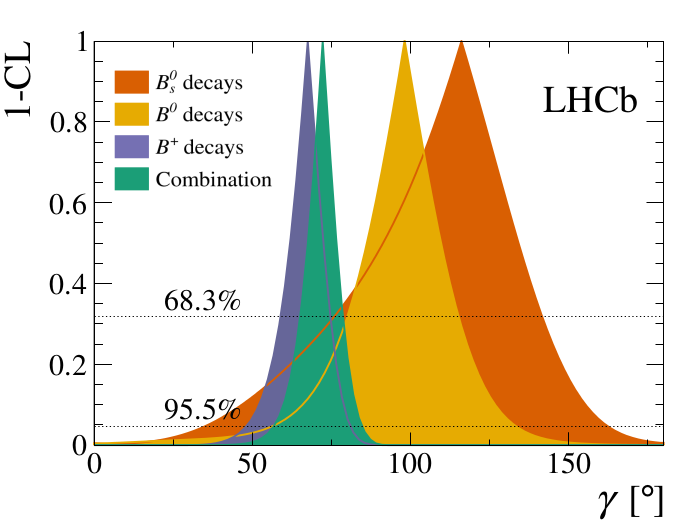}
\caption{1-CL curves for the combination of the $\gamma$ measurements and the contributions from the individual methods.}
\label{fig:gamma}
\end{figure}
All these different methods are used to perform measurements by LHCb using Run 1 data and the tree-level measurements have been combined to derive the best determination of $\gamma$.  To obtain the best precision, the hadronic parameters are also considered in the combination, for a total of 71 observables and 32 parameters. Both the frequentist and bayesian approaches have been used and they show a good agreement. Using the frequentist interpretation the combination is found to be: $\gamma = \left( 72.2^{+6.8}_{-7.3} \right)^{\circ}$~\cite{Aaij:2016kjh}, which is the most precise single-experiment measurement to date. The shapes of the 1-CL curves, split by the method, are shown in Figure \ref{fig:gamma}.

\subsection{Measurement of the CKM phase $\beta$}

\begin{figure}[h]
\centering
\includegraphics[height=2in]{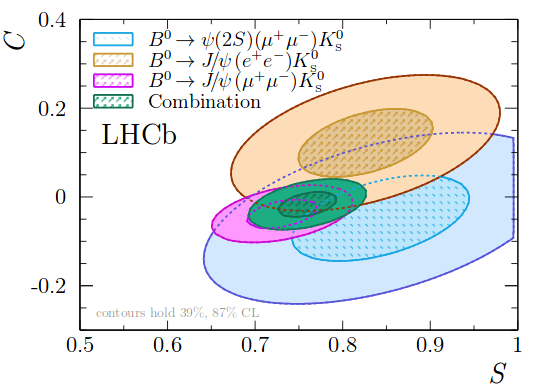}
\includegraphics[height=2in]{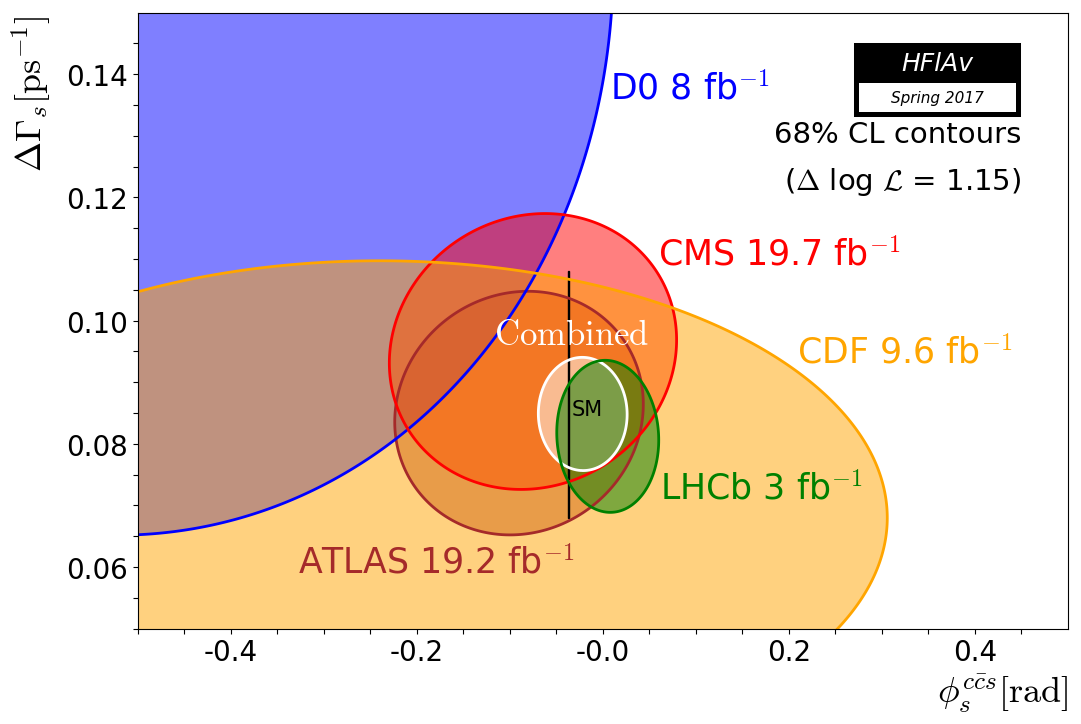}
\caption{Left: Two-dimensional likelihood scans for the combination of all $B^0 \to [c\overline{c}] K^0_s$ modes analysed by LHCb. Right: Two-dimensional likelihood scans of $\phi_s$ and $\Delta \Gamma_s$. The white contour represents the HFLAV combination to be compared to the SM prediction (black line).}
\label{fig:beta}
\end{figure}

The angle $\beta$ is related to CKM matrix elements through the relation $\beta \equiv \mathrm{arg} [ - (V_{cd}V_{cb^*})/(V_{td}V_{tb}^*) ]$ and it is sensitive to BSM effects in $C\!P$ violation in mixing and decay. The golden channel to determine this angle is $B^0 \to J/\psi K^0_s$. At LHCb $J/\psi \to \mu \mu$ decays are selected efficiently and the relatively high branching fraction of decays such as $B^0 \to J/\psi K^0_s$ gives large samples of $b \rightarrow c\overline{c} s$ transitions to be analysed. By measuring the flavour-tagged decay-time-dependent $C\!P$ asymmetry using the full Run I dataset, the LHCb collaboration measures~\cite{Aaij:2015vza} the following $C\!P$ violating parameters
\begin{equation}
  C = -0.038 \pm 0.032\,\mathrm{(stat.)} \pm 0.005 \,\mathrm{(syst.)}\,\, \mathrm{and}\,\, S \equiv \sin{(2\beta)} = 0.731 \pm 0.035\,\mathrm{(stat.)} \pm 0.020 \,\mathrm{(syst.)}\,.
\end{equation}
The largest systematic uncertainty on the measurement of $\sin(2\beta)$ arises from possible flavour asymmetries in the background candidate. The LHCb collaboration has also recently measured $S_f$ and $C_f$ using $B^0 \to J/\psi(e^+e^-) K^0_s$ and $B^0 \to \psi(2S)(\mu^+\mu^-) K^0_s$ decays~\cite{beta}. The individual LHCb measurements and their combination are shown in Figure~\ref{fig:beta} (left).
\subsection{Measurement of the CKM phase $\beta_s$}

The corresponding $C\!P$ observable in the $B^0_s$ system is called $\beta_s = -2\,\phi_s$ and it is related to CKM matrix elements through the relation $\phi_s^{\mathrm{SM}} \equiv -2\,\mathrm{arg} \left( - \frac{V_{ts}V_{tb^*}}{V_{cs}V_{cb}^*} \right)$. This angle has been extensively studied by ATLAS~\cite{Aad:2016tdj}, CMS~\cite{Khachatryan:2015nza} and LHCb~\cite{Aaij:2014zsa} collaborations using the golden channel $B^0_s \rightarrow J/\psi \phi$. This decay is a pseudo-scalar to vector-vector decay, so angular momentum conservation implies that the final state is an admixture of $C\!P$-even and $C\!P$-odd components. By performing a tagged time-dependent angular analysis, it is possible to statistically disentangle the different $C\!P$ eigenstates by the differential decay rate for $B^0_s$ and $\overline{B}^0_s$ mesons produced as flavour eigenstates at $t=0$. The LHCb collaboration exploited also other decay modes to fully investigate the potentials of Run 1 data. Different measurements have been performed using various $B^0_s$ finale states like $J/\psi \pi^+ \pi^-$~\cite{Aaij:2014dka}, $\psi (2S) \phi $~\cite{Aaij:2016ohx}, $D_s^+ D_s^-$~\cite{Aaij:2014ywt} and $J/\psi K^+K^-$~\cite{Aaij:2017zgz} for the $ K^+K^-$ invariant mass region above $1.05$ GeV. The HFLAV average including all the above mentioned results is $\phi_s = -21 \pm 31$ mrad~\cite{Amhis:2016xyh}. The individual measurements and their HFLAV combination are shown in Figure~\ref{fig:beta} (right). Despite the impressive progress since the initial measurements performed by CDF and D0, the uncertainty is still much larger than the indirect determination from global fits: $\phi_s^{\mathrm{SM}} = -36.5 ^{+1.3}_{-1.2}$ mrad~\cite{Charles:2015gya}.

\subsection{Time-dependent $\mathbf{\textit{CP}}$ violation in $B^0_s \to K^+K^-$ and $B^0 \to \pi^+\pi^-$}
The study of $C\!P$ violation in charmless charged two-body decays of neutral \textit{B} mesons is a good probe of the SM predictions. Moreover, precise measurements in this sector are important to constrain hadronic factors that cannot be accurately calculated from quantum chromodynamics at present. The LHCb collaboration performed a preliminary measurement of time-dependent $C\!P$-violating asymmetries in $B^0_s \to K^+K^-$ and $B^0 \to \pi^+\pi^-$ decays~\cite{BKK} using Run 1 data.
By means of a time-dependent analysis, where initial $B^0_{(s)}$ flavour is identified through a flavour-tagging algorithm calibrated using flavour-specific $B^0 \to K^+ \pi^-$ events, the parameters for $B^0_s \to K^+ K^-$ decay are found to be
\begin{equation}
  C_{KK} = 0.24 \pm 0.06\,\mathrm{(stat.)} \pm 0.02 \,\mathrm{(syst.)}\,\, \mathrm{and}\,\, S_{KK} = 0.22 \pm 0.06\,\mathrm{(stat.)} \pm 0.02 \,\mathrm{(syst.)}\,.
\end{equation}
The results for the $B^0 \to \pi^+\pi^-$ decay are
\begin{equation}
  C_{\pi\pi} = -0.24 \pm 0.07\,\mathrm{(stat.)} \pm 0.01 \,\mathrm{(syst.)} \,\, \mathrm{and}\,\, S_{\pi\pi} = -0.68 \pm 0.06\,\mathrm{(stat.)} \pm 0.01 \,\mathrm{(syst.)}\,.
\end{equation}
The value of $A^{\Delta \Gamma}_{KK}$ is found to be $-0.75 \pm 0.07\,\mathrm{(stat.)} \pm 0.11 \,\mathrm{(syst.)}$. Only one of the tagging algorithms available at LHCb, namely the so called opposite side taggers, has been used for this preliminary result, thus improvements in the tagging power are expected in the future. These results are compatible with SM predictions and are important inputs in the determination of the unitarity triangle angle $\gamma$ and $\beta_s$ using decays affected by penguin processes. 

\subsection{Time-dependent $\mathbf{\textit{CP}}$ violation in $B^0 \to D^+D^-$}\label{subsec:an3}
The $C\!P$ asymmetry obtained using $B^0 \to D^+D^-$ decays
\begin{equation}
\mathcal{A}_{C\!P}(t)=\frac{\Gamma(\overline{B}^0(t)\rightarrow D^+D^-)-\Gamma(B^0(t)\rightarrow D^+D^-)}{\Gamma(\overline{B}^0(t)\rightarrow D^+D^-)+\Gamma(B^0(t)\rightarrow D^+D^-)} = S \sin{(\Delta m_d,t)} - C \cos{(\Delta m_d t)}
\end{equation}
gives complementary information on $\sin{(2\beta)}$ through $b \rightarrow c\overline{c} d$ transitions. Indeed, this $C\!P$ asymmetry is sensitive to $\beta$ at tree level in the SM. However, higher order diagrams, such as penguin diagrams, can lead to a difference in the $C\!P$ violating phase, $\Delta \phi$, with respect to the parameter $\beta$ measured using $b \rightarrow c\overline{c} s$ transitions, such that $S/\sqrt{1-C^2}=-\sin{(2\beta - \Delta \phi)}$. Thus, a comparison with the $\beta$ measurement obtained using $B^0 \rightarrow J/\psi K^0_S$ decays constrains higher order contributions in $B \to DD$ decays that cannot be predicted by theory.\\
The LHCb collaboration performed a measurement of the $B^0 \to D^+D^-$ decay rate as a function of the $B^0$ decay
time~\cite{Aaij:2016yip}. The production flavour of the $B^0$ is determined by means of flavour tagging algorithms. The mistag rate ($\omega$) for the flavour tagging is
calibrated by measuring the $B^0$ oscillation amplitude in flavour specific $B^0 \to D_s^+ D^-$ decays. The tagging power in the $B^0 \to D^+D^-$ analysis is $\varepsilon_{\mathrm{eff}} \equiv \varepsilon(1-2\omega)^2= (8.1 \pm 0.6) \%$, where $\varepsilon$ is the fraction of events with a tagging decision.  This is the highest tagging power achieved at LHCb to date. The measured $S$ and $C$ parameters are
\begin{equation}
 S=-0.54^{+0.17}_{-0.16}\mathrm{(stat)}\pm 0.05 \mathrm{(syst)} \,\,\mathrm{and}\,\, C=-0.26^{+0.18}_{-0.17}\mathrm{(stat)}\pm 0.02 \mathrm{(syst)} \,.
\end{equation}
The phase shift derived from the LHCb measurement is $\Delta \phi = (0.16^{+0.19}_{-0.21}) \mathrm{rad}$, which rules out large penguin contributions.

\subsection{Time-dependent $\mathbf{\textit{CP}}$ violation in $B^0_s \to D_s^\mp K^\pm$}\label{subsec:an3}

Due  to  the  interference  between  mixing  and  decay  amplitudes, the $C\!P$ violating observables in the $B^0_s \to D_s^\mp K^\pm$ decays are functions of a combination of $\gamma$ and $\beta_s$, namely $\gamma - 2\beta_s$. By using an independent measurement on $\beta_s$ is thus possible to determine an independent value of $\gamma$. A comparison between the value of $\gamma$ measured from tree-level processes, as mentioned in Section \ref{subsec:gamma}, with the value of $\gamma$ and other unitarity triangle parameters measured in loop-level processes provides a powerful consistency check of the SM. In particular, in this case the amplitude ratio between the dominant tree level process and the subleading process is approximately 0.4. Moreover, the
decay-width difference in the $B^0_s$ system, $\Delta \Gamma_s$, is nonzero, which, compared to the $B^0$ case, gives access to two additional $C\!P$-violating observables which are coefficients of in time-dependent hyperbolical terms. Thus a good sensitivity on $\gamma$ is expected. The results found by the LHCb collaboration~\cite{BDsK} analysing Run 1 data are summarise in Table \ref{tab:table1}. Assuming the measured value of $\phi_s$ equal to $-2\beta_s$ as external input the value of $\gamma$ is determined to be $\gamma=(127^{+17}_{-22})^{\circ}$. The quoted uncertainties are
the sum in quadrature of the statistical and systematic components.  A compatibility of $2.15\, \sigma$ is found between the presented measurement and the LHCb $\gamma$ average presented in Section \ref{subsec:gamma}.

\begin{table}[h]
\begin{center}
\begin{tabular}{ll}  
\hline
    $C_{f}$ & = 0.735 $\pm$ 0.142 (stat) $\pm$ 0.048 (syst) \\ 
    $S_{f}$ & = -0.518 $\pm$ 0.202 (stat) $\pm$ 0.073 (syst) \\
    $S_{\overline{f}}$ & = -0.496 $\pm$ 0.197 (stat) $\pm$ 0.071 (syst) \\ 
    $A^{\Delta\Gamma}_{f}$ & = 0.395 $\pm$ 0.277 (stat) $\pm$ 0.122 (syst) \\
    $A^{\Delta\Gamma}_{\overline{f}}$ & = 0.314 $\pm$ 0.274 (stat) $\pm$ 0.107 (syst) \\ 

\hline
\end{tabular}
\caption{Time-dependent $C\!P$ violating observables in $B^0_s \to D_s^\mp K^\pm$ decays. The second and the third uncertainties are statistical and systematical respectively.}
\label{tab:table1}
\end{center}
\end{table}

\section{Measurement of $C\!P$ violation in baryons}\label{subsec:an5}

The vast majority of $C\!P$ violation measurements in beauty decays have focused on $B$ mesons due to the rich phenomenology available in meson mixing and the large datasets collected by the $B$-factories. However, the LHC copiously produces beauty also in the form of $\Lambda^0_b$ baryons, which gives large samples of $\Lambda^0_b$ baryons for $C\!P$ measurements. In the SM, sizeable $C\!P$  asymmetries  are  predicted  to  occur  in  decays  of  beauty  baryons  to
final states that contain no charmed hadrons. LHCb has reported the first measurement of $C\!P$ violation in the baryon sector using $\Lambda_b \to p\, \pi^- \pi^+ \pi^-$ decays~\cite{Aaij:2016cla}. $C\!P$ violation arises from interference between tree and higher order decay amplitudes. Since the two amplitudes are of similar size, large interference effects are possible. The search for a $C\!P$ violating asymmetry in these decays cannot rely on a comparison of $\Lambda^0_b$ and $\overline{\Lambda}^0_b$ decay rates, due to the poorly known $\Lambda^0_b - \overline{\Lambda}^0_b$ production asymmetry in $pp$ collisions. In order to observe $C\!P$ violation the following $C\!P$-odd observables have been defined $C_{\widehat{T}}=\overrightarrow{p}_p\cdot(\overrightarrow{p}_{h_1^-}\times \overrightarrow{p}_{h_2^+})$ for $\Lambda_b$ and $\overline{C}_{\widehat{T}} = \overrightarrow{p}_{\overline{p}}\cdot(\overrightarrow{p}_{h_1^+} \times \overrightarrow{p}_{h_2^-})$ for $\overline{\Lambda}^0_b$, where $\overrightarrow{p}$ stands for momentum and $h_1 h_2$ are the fastest $\pi^-$ in the $\Lambda^0_b$ reference frame and the single $\pi^+$, respectively. From these observables it is possible to build the following asymmetries, 
$A= [N(C_{\widehat{T}}>0)-N(C_{\widehat{T}}<0)]/[N(C_{\widehat{T}}>0)+N(C_{\widehat{T}}<0)]$ and $\overline{A}=[\overline{N}(-\overline{C}_{\widehat{T}}>0)-\overline{N}(-\overline{C}_{\widehat{T}}<0)]/[\overline{N}(-\overline{C}_{\widehat{T}}>0)+\overline{N}(-\overline{C}_{\widehat{T}}<0)]$, where $N$ and $\overline{N}$ are the number of $\Lambda^0_b$ and $\overline{\Lambda}^0_b$, respectively. From these asymmetries the $C\!P$- and $P$-violating observables, $a_{C\!P}$ and $a_P$, are constructed. They are defined as
$a_{C\!P}=\frac{1}{2}(A-\overline{A})$ and $a_P=\frac{1}{2}(A+\overline{A})$. Analysing the data collected by the LHCb detector in Run 1, with $\sim 6650$ reconstructed $\Lambda^0_b \to p \pi^- \pi^+ \pi^-$ signal decays, it was possible to find the first evidence of $C\!P$ violation in baryon decays, namely $a_{C\!P} \neq 0$ at 3.3 $\sigma$. This result is obtained from the combination of measurements of the asymmetry in bins of the 4-body phase space. This represents an important test of the validity of the SM in the baryon sector. No deviation from parity conservation was found.

\section{Measurement of $C\!P$ violation in charm}\label{subsec:an5}

After the first experimental observations of the slow
mixing rate of the $D^0-\overline{D}^0$ flavour oscillations~\cite{Aaij:2012nva}, the level of attention on charm physics increased significantly, thanks to the full range of probes that has been disclosed, entirely
complementary to the $B$ and $K$ mesons, for mixing and $C\!P$ violation. Indeed the charm quark is the
only up-type quark that manifests flavour oscillations and the only way to probe $C\!P$ violation with up-type quarks. Moreover, only with the advent of LHC has it been possible to collect huge and very clean samples of $D$ meson decays, several orders of magnitude larger in
size than in the past, allowing for the first time to approach the small SM expectations for $C\!P$-violation, which is below the $10^{-3}$ level~\cite{Bhattacharya:2012ah}. 

\begin{figure}[h]
\centering
\includegraphics[height=2in]{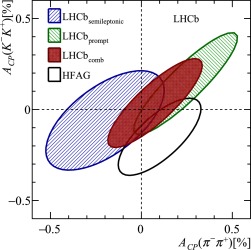}
\includegraphics[height=2.5in]{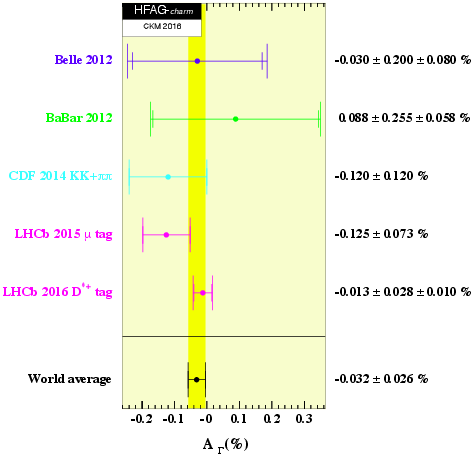}
\caption{Left: $C\!P$ violating asymmetries in $D^0 \to K^+K^-, \pi^+\pi^-$ decays using semileptonic $B$ decays-tagged and promptly-produced $D^{*+}$ tagged samples and their combination. Right: HFLAV combination of the parameter $A_{\Gamma}$.}
\label{fig:charm}
\end{figure}

Direct $C\!P$ violation can be experimentally determined by measuring the raw asymmetries in the decay of $D^0$ mesons to $C\!P$ eigenstates, $f$, for example $D^0 \to K^+K^-, \pi^+\pi^-$
\begin{equation}
A_{raw} = \frac{N(D^0 \to f) - N(\overline{D}^0 \to f)}{N(D^0 \to f) + N(\overline{D}^0 \to f)}\,.
\end{equation}
The $C\!P$ asymmetry is then determined taking into account the production, $A_{P}$, and the detection $A_{D}$ asymmetries, $A_{raw} \approx A_{CP} + A_{P} + A_{D}$.
The most recent LHCb measurements are~\cite{Aaij:2016dfb} $A_{C\!P} (K^+K^-) = (0.04 \pm 0.12 \mathrm{(stat)} \pm 0.10 \mathrm{(syst)})\%$ and $A_{C\!P} (\pi^+\pi^-) = (0.07 \pm 0.14 \mathrm{(stat)} \pm 0.11 \mathrm{(syst)})\%$, which are a combination of a semileptonic $B$ decays-tagged and promptly-produced $D^{*+}$ tagged samples. Figure~\ref{fig:charm} (left) shows the results for the two $C\!P$ asymmetries. LHCb also recently studied $C\!P$ asymmetries in $D^+_{(s)} \to \eta' (\to \pi^+ \pi^- \gamma) \pi^+$~\cite{Aaij:2017eux}, which was never measured before at a hadron collider due to the difficulty of accurately modelling the large background and obtained the most precise measurement to date in this decay. LHCb also studied $C\!P$ asymmetries in $D^0 \to \pi^+\pi^-\pi^+\pi^-$~\cite{Aaij:2016nki} decays, which is the first time that such a study has been performed for a 4-body decay, for which the rich resonant structure allows to search for local enhancement of the $C\!P$ asymmetries. These results are now approaching the per-mille level of uncertainty, but so far they are consistent with $C\!P$ conservation.

Neutral D mesons oscillations are dominated by long-distance effects and the mixing parameters are small, as are the the expected $C\!P$-violating effects. Indirect $C\!P$ violation can be determined by measuring the decay-time dependent $C\!P$ asymmetry of $D^0$ decays to $C\!P$ eigenstates, $f$,
\begin{equation}
A_{C\!P}(t) = \frac{\Gamma(D^0(t) \to f) - \Gamma(\overline{D}^0 \to f)}{\Gamma(D^0 \to f) + \Gamma(\overline{D}^0 \to f)} \simeq a_{dir}^{f} - A_{\Gamma} \frac{t}{\tau_D}\;\;\;\;\; \mathrm{, where}\;\;\; A_{\Gamma} \equiv \frac{\widehat{\Gamma}_{D^0 \to f}-\widehat{\Gamma}_{\overline{D}^0 \to f}}{\widehat{\Gamma}_{D^0 \to f}+\widehat{\Gamma}_{\overline{D}^0 \to f}}\,.
\end{equation}
Here $A_{\Gamma}$ is the asymmetry between the $D^0$ and $\overline{D}^0$ effective decay widths and it is the parameter sensitive to $C\!P$ violation in the mixing and in the interference between mixing and decay.
Within the SM, the indirect determination of this parameter is $A_{\Gamma}<\mathcal{O}(10^{-3})$~\cite{Amhis:2016xyh}. LHCb recently measured it using a promptly-produced $D^{*+}$ tagged $D^0$ sample. The combination of $f=KK,\pi\pi$ measurements in Run I is $A_{\Gamma} = (0.13 \pm 0.28 \mathrm{(stat)} \pm 0.10 \mathrm{(syst)})\times10^{-3}$~\cite{Aaij:2017idz}. The result is consistent with the hypothesis of $C\!P$ conservation and represents the most precise measurement of $C\!P$ violation in the charm system ever made. In Figure~\ref{fig:charm} (right) the HFLAV~\cite{Amhis:2016xyh} combination of different $A_{\Gamma}$ measurements performed so far is shown.

\section{Conclusions}

Recent LHCb results of $C\!P$ violation in the beauty and charm hadron systems have been presented. All measurements are in good agreement with SM predictions. The selection of $C\!P$ violation measurements presented here is not a complete list of the analysis performed by the ATLAS, CMS and LHCb collaborations. Updated results based on the Run 2 data, collected in 2015 and 2016, are expected soon and will enable even stronger constraints to be made on potential BSM physics contributions.

%%  if necessary
% \Acknowledgements
% I am grateful to XYZ for fruitful discussions.

\end{document}